\documentclass{iau} 
\usepackage{graphicx}
\usepackage{gensymb}
\usepackage{natbib}

\newcommand{\he}{He\,{\sc i} 10830\,\AA}
\newcommand{\kms}{km\,s$^{-1}$}

\title[]
{Fast downflows in a chromospheric filament}

\author[Sowmya et al.]
{K. Sowmya$^1$, A. Lagg$^1$,
S. K. Solanki$^{1,2}$ and J. S. Castellanos Dur\'an$^1$}

\affiliation{$^1$Max-Planck-Institut f\"ur Sonnensystemforschung\\
Justus-von-Liebig-Weg 3, 37077 G\"ottingen, Germany \\ email: {\tt krishnamurthy@mps.mpg.de} \\[\affilskip]
$^2$School of Space Research, Kyung Hee University,
YongIn, Gyeonggi 446--701, Korea}

\pubyear{2019}
\volume{354}
\jname{Solar and Stellar Magnetic Fields: Origins and Manifestations}
\editors{A. Kosovichev, K. G. Strassmeier \& M. Jardine}
\begin{document}

\maketitle

\begin{abstract}
An active region filament in the upper chromosphere is studied using spectropolarimetric
data in \he\ from the GREGOR telescope. A Milne-Eddingon based inversion of the
Unno-Rachkovsky equations is used to retrieve the velocity and the magnetic field
vector of the region. The plasma velocity reaches supersonic values closer to the feet of
the filament barbs and coexist with a slow velocity component. Such
supersonic velocities result from the acceleration of the plasma as it
drains from the filament spine through the barbs. The line-of-sight magnetic fields
have strengths below 200\,G in the filament spine and in the filament barbs where
fast downflows are located, their strengths range between 100 - 700\,G.


\end{abstract}

\firstsection 
\section{Introduction}
Solar filaments are dense threads suspended in the chromosphere and
held in place by strong magnetic fields in the solar atmosphere. A filament
appears dark when observed on the disk with the chromospheric lines such as the
\he\ triplet, as the gas inside it is cooler than the photospheric plasma below.
Its structure resembles that of a centipede with the elongated body called
`spine' and the many legs called `barbs' which are rooted in the solar surface.

High speed downflows have been observed in the filament barbs (see e.g.,
\citealt{Joshietal13,Sassoetal11,Sassoetal14})
where the magnetic field is more vertical than in the filament itself.
Recently, \citet{Diazbasoetal19a,Diazbasoetal19b} explored the
dynamic and magnetic properties of an active region (AR) filament observed with
the ground based 1.5\,m GREGOR telescope \citep{Schmidtetal12},
with an emphasis on understanding its magnetic topology. An overview about
chromospheric filaments can be found in these papers and
the references therein. We study the same AR filament that was studied by
these authors but focus on the line-of-sight (LOS) velocity distribution
and magnetic field strength at selected locations in the barbs.

\section{Observations}
\begin{figure}[htbp]
\begin{center}
\includegraphics[scale=0.55,trim=7cm 3.5cm 7.cm 4cm,clip]{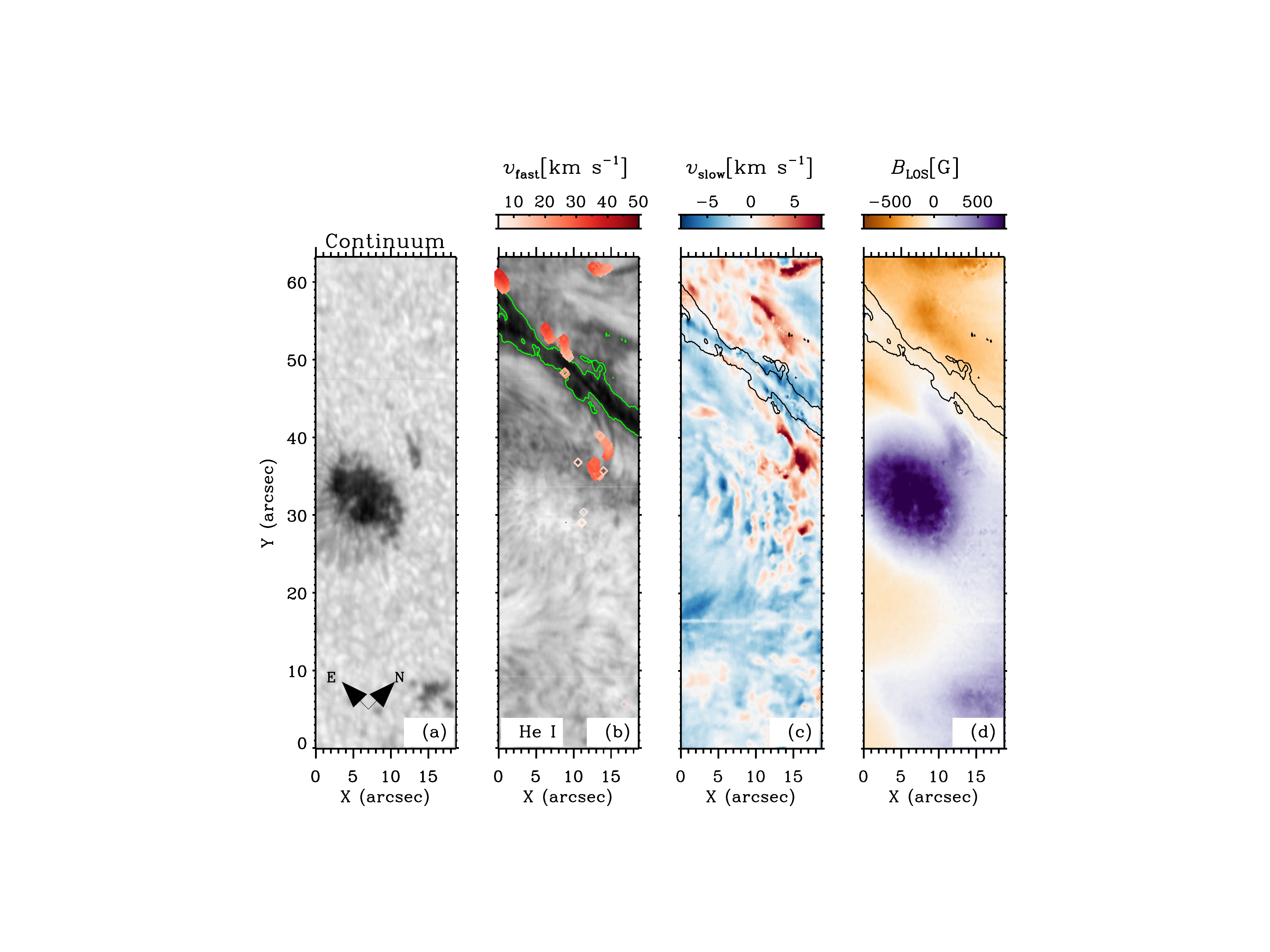}
\end{center}
\caption{Panel (a): continuum intensity image with the arrows showing the
directions of solar north and east. Panel (b): normalized intensity
in the He\,{\sc i} line at 10830.34\,\AA{} with the overplotted colored symbols
representing the LOS velocity of the fast component. Panel (c): LOS velocity
map for the slow component. Panel (d): LOS magnetic field strength map for
the slow component saturated at $\pm800$\,G. The green and black contours at
0.6\,$I_c$ mark the filament spine.}
\label{fig:maps}
\end{figure}
The spectropolarimetric raster scan of the filament associated with the AR 12087
was recorded on 17 June 2014 using the GREGOR Infrared Spectrograph
\citep[GRIS;][]{Colladosetal12}. The scan (ID: 17jun14.005)
lasted approximately 15 minutes (09:13 UT - 09:27 UT), covering about $19''$ in the
scan direction (x-axis) and $63''$ along the slit (y-axis), with a step size of
$0.126''$ and a pixel size of $0.135''$ along the slit. The coordinates of the
center of the field-of-view (FOV) correspond to a heliocentric angle of $23\degree$
($\mu = {\rm cos}\theta = 0.92$, $x=+180''$ and $y=-320''$). The data was reduced
with the standard GRIS data reduction software. The image resolution
determined by averaging the power spectrum along the slit direction is $0.5''$.
The observed spectral window of about 18\,\AA{} includes the chromospheric
\he{} triplet, photospheric Si\,{\sc i} 10827\,\AA{} and Ca\,{\sc i} 10839\,\AA{}
lines and a telluric blend at 10832\,\AA{}.
The average noise in the Stokes
parameters is $1.2\times10^{-3}I_c$, where $I_c$ is the continuum intensity.
Panels (a) and (b) in Fig.~\ref{fig:maps} show the intensity images in the continuum redward of
the Si line and at He\,{\sc i} 10830.34\,\AA{}. In the observed FOV, a part of
the filament spine and a few barbs are visible in He\,{\sc i}.
\section{Method}
We invert the spectropolarimetric data using the {\sc HeLIx$^+$} inversion code
assuming Milne-Eddington atmosphere and taking into account the Zeeman effect
in the incomplete Paschen--Back regime \citep{Laggetal04,Laggetal09}.
The Stokes profiles at some locations along the filament barbs show the
presence of a second velocity component which is strongly redshifted.
A single component inversion does not yield a good fit to such profiles (see
panels (a) and (b) in Fig.~\ref{fig:prof}). Therefore, we invert the He\,{\sc I}
triplet by using two atmospheric components which are mixed through a filling
factor. We call the two components `slow' ($\pm10$\,\kms) and `fast'
(downflows up to 50\,\kms).
From such inversions, we retrieve the LOS velocity and the magnetic field
vector for the two atmospheric components.

\section{Results}
The maps of intensity, LOS velocity and LOS magnetic field strength for
the slow component are shown in Fig.~\ref{fig:maps}. The green and black
contours at 0.6\,$I_c$ identify the filament spine. The slow component
is nearly at rest and its LOS velocity (see panel (c)) follows a Maxwellian
distribution. The LOS magnetic field strength map (see panel (d)) is
saturated at $\pm800$\,G. Chrompsheric magnetic
fields stronger than 800\,G are clearly present in the sunspot. The filament
is located above the polarity inversion line where the magnetic field
is changing sign. Inside the filament spine, the LOS component of the
magnetic field is weaker than 200\,G, in agreement with the previous reports
\citep[see e.g.,][]{Kuckeinetal09}.

The locations where a fast component with a filling factor greater
than 20\,\%\ and a signal in any Stokes polarization parameter higher
than three times the noise (within $\pm0.35$\,\AA{} from it's wavelength position)
is present are marked by red symbols in
the He\,{\sc i} intensity image in Fig.~\ref{fig:maps}. The colorbar
indicates the velocity of the fast component. We found no location
where more than two atmospheric components are present. It is apparent that
the fast component lies along the filament barbs. The velocity seems
to be increasing down the barb reaching supersonic values (higher
than the local sound speed of $\sim$10\,\kms) closer to the footpoints.
This increase can be explained by the gravitational acceleration of the plasma
from the spine as it drains along the barbs.
\begin{figure}[ht]
\begin{center}
\includegraphics[scale=0.43,trim=3.cm 9.5cm 2.cm 5cm,clip]{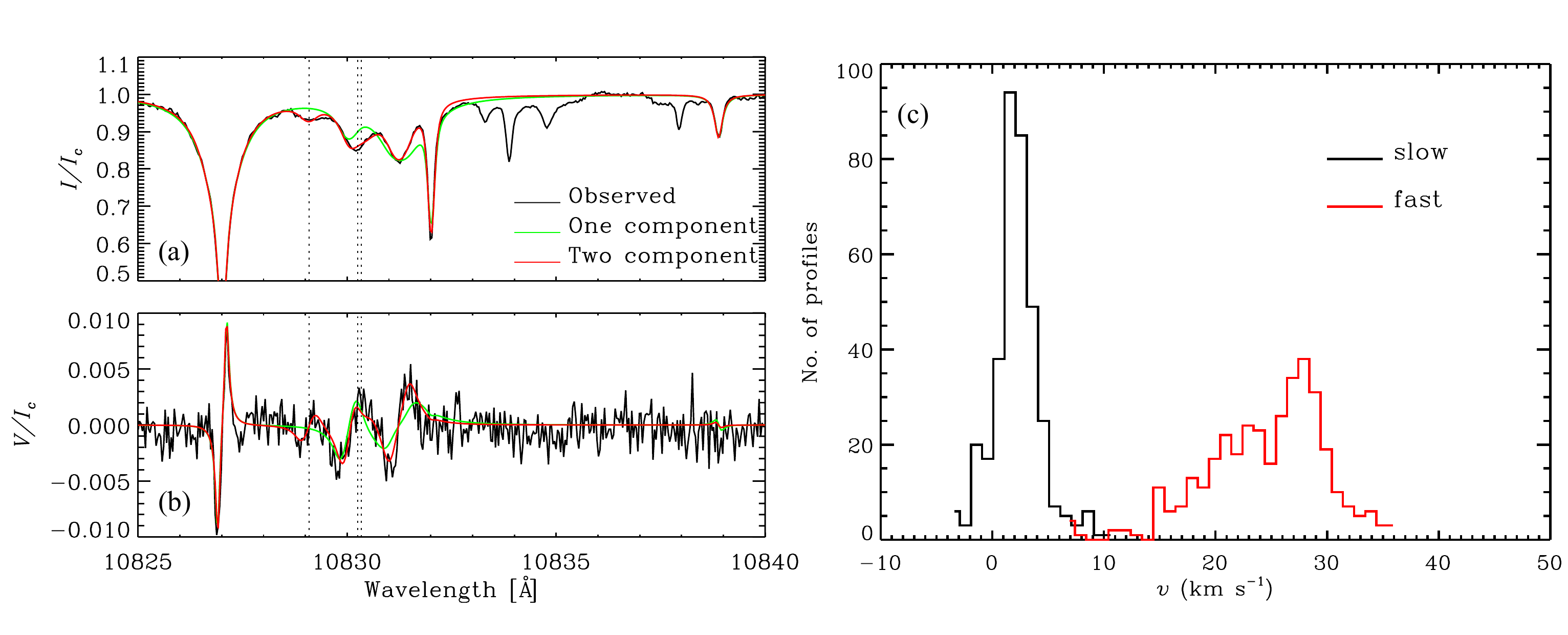}
\end{center}
\caption{Panel (a): observed intensity profile (black) at a location with
the coordinates $x=8.7''$ and $y=52.5''$ (see Fig.~\ref{fig:maps}), a fit to the
observed profile obtained from one component inversion (green) and a fit from
two component inversion (red). Panel (b): same as panel (a) but for
Stokes $V$. Panel (c): histograms of the LOS velocity at locations where
both the slow (black) and the fast (red) components are present.}
\label{fig:prof}
\end{figure}

Panels (a) and (b) in Fig.~\ref{fig:prof} show Stokes
$I$ and $V$ profiles at one of the locations corresponding to $x=8.7''$
and $y=52.5''$, where the fast component has a velocity of $\sim$30\,\kms.
Stokes $Q$ and $U$ signals are close to the noise level and hence we do
not show them here.
The observed profiles are shown in black, the fits from one component and
two component inversions are shown in green and red, respectively. Clearly,
two components are needed to properly model such anomalous profiles. We
identified 360 locations where a fast component is present simultaneously
with the slow component. Fig.~\ref{fig:prof}(c) shows the velocity distribution
for the fast (red) and slow (black) components at those locations. Most of
the downflow velocities range between 15 - 30 \kms{} and in some extreme cases
they reach 35 \kms{} or more. Along the filament barbs, the fast downflow
component has LOS magnetic field strength ranging between 100 - 700\,G.

\section{Conclusions}
We studied an active region filament observed by the ground based GREGOR
telescope. Using two component Milne-Eddington inversions, we determined the
velocity and the magnetic field vector at chromospheric heights where He I forms.
We identified supersonic downflows, as high as 35\,\kms, coexisting with a slow
flow component. The high donwlfow velocities can be understood
with the acceleration of the plasma as it drains along the filament
barbs. The slow component could be originating from He\,{\sc i} layer
at a different height or could be due to straylight.
We find that the LOS component of the magnetic field is weaker
than 200\,G within the filament spine. In filament barbs where strong
downflows are found, the LOS magnetic field is as strong as 700\,G.

\begin{acknowledgements}
Marie Sk{\l}odowska-Curie grant agreement No. 797715. The 1.5-meter GREGOR 
solar telescope was built by a German consortium under the leadership of the
Leibniz Institut f\"ur Sonnenphysik in Freiburg with the Leibniz Institut
f\"ur Astrophysik Potsdam, the Institut f\"ur Astrophysik G\"ottingen, and
the Max-Planck Institut f\"ur Sonnensystemforschung in G\"ottingen as partners,
and with contributions by the Instituto de Astrof\'isica de Canarias and the
Astronomical Institute of the Academy of Sciences of the Czech Republic.
The GRIS instrument was developed thanks to the support by the Spanish Ministry
of Economy and Competitiveness through the project AYA2010-18029 (Solar
Magnetism and Astrophysical Spectropolarimetry). This study
has made use of SAO/NASA Astrophysics Data System's bibliographic service
\end{acknowledgements}

\end{document}